# Suppression of Dyakonov-Perel Spin Relaxation in high mobility n-GaAs

R.I. Dzhioev, K.V. Kavokin, V.L. Korenev*, M.V. Lazarev, N.K. Poletaev, B.P. Zakharchenya,

*A.F. Ioffe Physical Technical Institute*, *St. Petersburg, 194021 Russia*

E.A. Stinaff, D. Gammon, A.S. Bracker, M.E. Ware

*Naval Research Laboratory, Washington DC 20375, USA*

We report a large and unexpected suppression of the free electron spin relaxation in lightly-doped n-GaAs bulk crystals. The spin relaxation rate shows weak mobility dependence and saturates at a level 30 times less then that predicted by the Dyakonov-Perel theory. The dynamics of the spin-orbit field differs substantially from the usual scheme: although all the experimental data can be self-consistently interpreted as a precessional spin relaxation induced by a random spin-orbit field, the correlation time of this random field, surprisingly, is much shorter than, and is independent of, the momentum relaxation time determined from transport measurements. Understanding of this phenomenon could lead to high temperature engineering of the electron spin memory.

*Corresponding author: korenev@orient.ioffe.rssi.ru

Electrons in n-doped semiconductors demonstrate long spin relaxation times $(\tau_s)$ at liquid-helium temperatures [1,2,3,4], because spin-orbit channels of spin relaxation (Dyakonov-Perel and Elliot-Yaffet mechanisms [5]) are frozen out. When temperature increases, the electrons bound to shallow donors dissociate to the conduction band, where $\tau_s$ is much shorter because of the spin-orbit interaction. This fact imposes serious limitations on high-temperature spintronics [6]. For example, spin-polarized charge carriers, optically or electrically injected into the semiconductor, considerably depolarize at elevated temperatures [5,7]. If the non-equilibrium spin of electrons is to be used for applications, the nature of spin relaxation at relatively high temperatures should first be understood. At T>50 K, the Dyakonov-Perel (DP) [8] mechanism is believed to dominate spin relaxation in p-GaAs [5] and n-GaAs [2,9]. This mechanism can be interpreted in terms of an effective magnetic field of spin-orbit interaction, which acts upon the electron spin in semiconductor crystals without inversion symmetry. The value and direction of this "spin-orbit" field are determined by the value of the electron **k**-vector and its direction with respect to the crystal axes. The standard DP approach [8] assumes that the frequent collisions with impurities or phonons, which change the electron wave-vector, make the spin-orbit field fluctuate, resulting in a dynamical suppression of spin relaxation. As a result the spin relaxation rate is described by the motional-narrowing formula: $\tau_s^{-1} = \langle \Omega^2 \rangle \tau_*$, where $\tau_*$ is a correlation time proportional to the momentum relaxation time $\tau_p$. Faster scattering leads to slower spin relaxation [5, 8]. Therefore, the spin relaxation rate increases with the electron mobility, and the DP mechanism should dominate in "clean" samples with low doping.

In this work we show that, against expectations, the standard DP mechanism is suppressed in "clean", high-mobility GaAs samples. Measurements on a set of samples with the electron mobility $\mu_n$ ranging from 5,000 to 200,000 $cm^2/V*s$ show that the spin relaxation rate first grows with mobility, but then, at $\mu_n$>20,000 $cm^2/V*s$, saturates at the level of (1.5±0.5) $ns^{-1}$ at T=77K, where electrons obey non-degenerate (Boltzmann) statistics. This finding is in sharp contradiction with the DP theory, which predicts further increase of the relaxation rate. The fact that in high-mobility samples relaxation is 30 times *slower* than the DP prediction might be interpreted as a suppression of the DP mechanism. The suppression persists up to the room temperature. Further experiments in high-mobility pure samples in a longitudinal magnetic field show that the real dynamical picture differs from the usual DP scheme: although all the experimental data can be self-consistently interpreted as a precessional spin relaxation induced by a random spin-orbit (Dresselhaus) field, the correlation time of this random field, surprisingly, is much shorter than the momentum relaxation time determined from transport measurements. It weakly depends on mobility and is approximately equal to 300 fs for most pure samples. It means that electron spin relaxation in high-mobility n-type semiconductors decouples from the drift transport of charge.

We have measured a large number of samples grown with different technologies. Both mobility and concentration were measured using the Hall effect in the dark. In one set of samples, grown with molecular beam epitaxy (MBE), 2-7.5µm thick GaAs layers were sandwiched between AlGaAs barriers. Concentrations of itinerant electrons were from $10^{14}$ to $6*10^{16}$ $cm^{-3}$ (mobilities $\mu_n$ from *140,000* to *5,000 $cm^2/V*s$*, respectively). The second group of samples, grown by liquid-phase epitaxy (LPE), had 20-30µm thick GaAs layers with electron concentrations from $4.5*10^{14}$ to $3*10^{16}$ $cm^{-3}$ and mobilities $\mu_n$ from *80,000* to *11,000 $cm^2/V*s$*. In the third group, grown by gas-phase epitaxy (GPE), GaAs layers were 30-40µm thick with electron concentrations from $2*10^{13}$ to $3*10^{15}$ $cm^{-3}$ and mobilities $\mu_n$ from *200,000* to *20,000 $cm^2/V*s$*. All the samples were grown in the [001] direction.

The samples were placed in a liquid-nitrogen cryostat and pumped by a tunable Ti-sapphire laser, with the circular polarization of light being alternated in sign at a frequency of 27 kHz with a photoelastic quartz modulator. This allowed us to eliminate the effect of lattice nuclear polarization on the optical orientation of the electrons [5]. The photoluminescence (PL) polarization was measured in the reflection geometry by a circular-polarization analyzer. A two-channel photon counting device synchronized with the quartz modulator provided measurement of the effective degree of circular polarization $\rho_c = (I_+^+ - I_+^-)/(I_+^+ + I_+^-)$, where $I_+^+$, $I_+^-$ are the intensities of the $\sigma^+$ PL component under $\sigma^+$ and $\sigma^-$ pumping, respectively.

The optical orientation method, as applied to measuring spin relaxation time in n-GaAs, is based on the following [5]. Circularly polarized light creates partially spin-polarized electrons. The simultaneously created holes rapidly lose their spin orientation. Their recombination with electrons of both spin directions accumulates the non-equilibrium mean spin of the electron ensemble. The experimentally measured circular polarization, $\rho_c$, is proportional to $S_z$, the projection of the electrons' average spin onto the excitation light beam direction. An external magnetic field, perpendicular to the pump- and PL direction (Voigt geometry), makes the spin rotate with the Larmor precession frequency $\omega = \mu_B g_e B/\hbar$ (where $g_e$ is the electron g-factor, and $\mu_B$ is the Bohr magneton). As a result of the precession, $S_z$ decreases with increasing magnetic field, and is observed as PL depolarization (Hanle effect) with Lorentzian lineshape and a halfwidth, $B_{1/2} = \hbar/\mu_B g_e T_s$. The inverse spin lifetime is $T_s^{-1} = \tau_s^{-1} + \tau_J^{-1}$. Specifically in n-type semiconductors, the lifetime $\tau_J$ of itinerant electrons depends on their equilibrium concentration $n=N_d-N_a$ (where $N_d$ and $N_a$ are concentrations of donors and acceptors, respectively) and on the pump density, $G$, as $\tau_J = n/G$ [5]. Therefore, in the limit of low pump density, $T_s$ is equal to the electron spin relaxation time $\tau_s$, whereas the degree $\rho_c \sim \tau_s/\tau_J$.

The spectra of the PL intensity (solid curve) and the polarization (circles) taken at 77K in zero magnetic field are shown in Fig.1a for a thin (less than electron spin diffusion length $L_s$ [1]), 2μm, MBE-grown GaAs layer ($N_d-N_a=3*10^{15}$ $cm^{-3}$, $\mu_n=29,000$ $cm^2/V*s$). The PL band has a pronounced short-wavelength wing due to recombination of free electrons and holes. The polarization is nearly constant over the PL spectrum. The Hanle curve, measured at the PL maximum under quasi-resonant excitation ($h\nu_{exc}=1.521$ $eV$) with pump density $W=1$ $W/cm^2$, is shown in Fig.1b. The Hanle curve remains the same through the entire PL spectrum and is fit well by a Lorentzian. Its half-width allows us to determine, in the weak-pump limit, the electron spin relaxation rate $1/\tau_s=(0.9\pm0.1)$ $ns^{-1}$.

In the thick (larger than $L_s$) samples the PL intensity spectrum (Fig.2a) is nearly the same as in the thin samples (Fig.1a), but the polarization decreases from high-energy (4.5 %) to low-energy (1.5 %) (circles in Fig.2a) because of electron spin diffusion [1, 5,10]. Nevertheless, the half-width of the Hanle curves measured in the high and low-energy tails of PL differs only by 30 %. The effect of spin diffusion is negligible in the low-energy ($h\nu_{det}=1.505$ $eV$) tail [1, 10] where the measurements on all thick samples were performed. An example of the Hanle effect at $W=1$ $W/cm^2$ is shown in Fig.2b. The pump dependence of $B_{1/2}$ yields $1/\tau_s=(1.0\pm0.1)$ $ns^{-1}$.

The dependence (circles) of the spin relaxation rate $1/\tau_s$ on the electron mobility is the main result at T=77 K (Fig.3). Within the range of mobility from $\mu_n=5,000$ $cm^2/V*s$ to $20,000$ $cm^2/V*s$ the rate $1/\tau_s$ increases with mobility, but then, at high-mobility, it saturates at the level of $(1.5\pm0.5)$ $ns^{-1}$. This result suggests also a weak dependence of the rate $1/\tau_s$ on concentration $n=N_d-N_a$ because the variation of mobility at a fixed temperature is the result of variation of the doping level. It agrees well with experiment (inset on Fig.3).

Our experimental results disagree with the standard theory of DP spin relaxation, which predicts proportionality between $1/\tau_s$ and the electron mobility. Indeed, the spin relaxation rate for the non-degenerate statistics of electrons [11] is given by the expression [5]

$$\frac{1}{\tau_s} = Q_1 \alpha^2 \tau_p \frac{(kT)^3}{E_g \hbar^2}$$

(1)

where the spin-orbit parameter $\alpha=0.07$, $E_g$ is the GaAs band gap. $Q_1$ is a number of the order of 1, depending on the scattering mechanism ($Q_1=3$ for scattering by polar optical phonons [12] and *1.5* for scattering by ionized impurities). Our results demonstrate a breakdown of Eqn. (1) in high-mobility samples. The theoretical dependence of $1/\tau_s$, according to Eq.1, is plotted in Fig.3 for $Q_1=3$ (upper straight line) and *1.5* (lower straight line). In low-mobility samples the momentum scattering is determined by ionized impurities, while in high-mobility samples it is governed by phonons. Therefore the theory predicts that the dependence of $1/\tau_s$ on mobility should lie in between the two straight lines, starting from the lower one at low mobilities, and getting closer to the higher one at high mobilities (the theoretical limit is about *260,000 cm²/V\*s* at 77 K). One can see that the theory disagrees with the experiment both qualitatively and quantitatively: for high-mobility samples the experimentally measured $1/\tau_s$ is more than an order of magnitude smaller than predicted by the DP theory and does not depend on mobility [13].

We are unlikely to understand this puzzling suppression result by searching for a *new* source of spin relaxation because an additional mechanism would only increase the $1/\tau_s$ rate in comparison with that given by the DP-mechanism alone. Therefore, the main question to be understood here is the reason for the giant suppression of DP spin relaxation. The experiments on optical orientation in longitudinal magnetic field (Faraday geometry) provide an unambiguous answer: the spin-orbit (Dresselhaus) field in high-mobility GaAs is averaged out through a *new* mechanism that is much more efficient than and has nothing to do with the momentum scattering averaging. It is well known [5] that the optical orientation of electrons in a longitudinal magnetic field gives the correlation time of the random field responsible for the spin relaxation. In the case of free electrons, the effect of the longitudinal magnetic field is associated with the cyclotron rotation of the electron momentum around **B** with the frequency, $\omega_c = eB/mc$. This rotation averages the momentum components perpendicular to **B**, thus diminishing the effective spin-orbit field. As a result, both $\tau_s$ and $\rho_c \sim \tau_s/\tau_J$ increase as a function of B. If **B** is directed along [100] and is not very strong ($\omega_c \tau_p < 1$), then the dependence $1/\tau_s(B)$ is parabolic [5]

$$\frac{1}{\tau_s(B)} = \frac{1}{\tau_s(0)}\left[1 - \Theta_2 (\omega_c \tau_p)^2\right]. \qquad (2)$$

Comparing Eq.2 with experiment, one can determine the momentum relaxation time. The parameter $\Theta_2 \approx 2$ [5] depends slightly on the specific scattering mechanism. The dependence of $\rho_c$ on the longitudinal magnetic field is shown by circles in Fig.2b, inset. The solid curve is a fit using Eq.(2) resulting in $\tau_p \approx 270$ *fs*. This is 20 times *shorter* than the momentum scattering time taken from the Hall mobility, $\tau_p = 5.4$ *ps* ($\mu_n = 140,000$ $cm^2/V*s$). Substitution of the value of $\tau_p$ obtained from the longitudinal-field measurements into the expression for $1/\tau_s$ (Eq.1) gives the following results. For impurity scattering, we get $1/\tau_s = 1$ $ns^{-1}$ (lower triangle in Fig.3), and for phonon scattering $1/\tau_s = 1.5$ $ns^{-1}$ (upper triangle in Fig.3). These values are in good agreement with the value $1/\tau_s = 1.1$ $ns^{-1}$, measured independently in Voigt geometry (Hanle effect, circle in Fig.3). In the same manner, using the longitudinal-field measurements and Eq.(1), we have plotted spin relaxation for several other samples with different Hall mobilities (triangles in Fig.3). The agreement with the spin relaxation rates taken from the Hanle effect (circles in Fig.3) is fairly good. We wish to stress that for all the samples for which the Faraday-geometry measurements were performed, the "momentum relaxation time" calculated using Eq.(2) (in fact, the *correlation time* of the spin-orbit field) was much shorter than $\tau_p$ taken from the Hall effect, and did not depend on mobility.

Temperature dependence of the spin relaxation rate $1/\tau_s$ gives further evidence of the failure of the classical DP approach in high mobility GaAs. The rate was measured (circles) on 7.5μ thick MBE-grown GaAs (n=9*10$^{13}$ cm$^{-3}$, *$\mu_n$=140,000 cm$^2$/V*s* at T=77 K) at the center of the PL line (Fig.4). One can see that the rate $1/\tau_s$ grows from 1 ns$^{-1}$ up to 20 ns$^{-1}$ in the temperature range 50-250 K. The solid line in Fig.4 is calculated according to Eq.(1) with the feature around 150 K being determined by the change of averaging mechanism from acoustic phonon and ionized impurity (T<150 K) to polar optical phonon scattering (T>150 K). Squares show the extension [14] of the DP scheme taking into account the non-elastic character of the optical phonon scattering. In spite of the correct trend there is quantitative disagreement between theory and experiment over the entire range of temperatures we have studied.

Our results indicate that fast dynamical averaging of the spin-orbit field (over the thermal distribution of **k**) gives a self-consistent and very precise description of spin relaxation in n-GaAs. *But, quite surprisingly, the correlation time $\tau_c$ of this field has nothing in common with the momentum scattering time $\tau_p$ in high-mobility samples.* At the first glance, electron-electron scattering can be invoked as a possible explanation since it suppresses spin relaxation but does not influence the mobility [15,16]. However, though effective in doped samples with low mobility, it plays no role in clean samples with high mobility and can be ruled out. Another possibility is that, unlike the Hall effect, the Hanle effect measures the spin memory of *localized* near-band-gap electrons. Localization would suppress the DP mechanism. However, localization is highly unlikely in high quality samples at elevated temperatures. Therefore, the data lead us to conclude that ballistically flying electrons experience a rapidly fluctuating ($\tau_c \approx 300$ fs) spin-orbit field. However, the origin of the rapid fluctuations remain unknown. We might imagine that the sign of the spin-orbit "constant" $\alpha$ fluctuates in space, inducing space fluctuations of the spin-orbit field. In this case the correlation time is determined by the time-of-flight of the electron over a typical fluctuation length. This mechanism has been suggested for Si/Ge quantum wells [17], where the fluctuations are governed by the local symmetry of interfaces. The origin of the fluctuations in bulk GaAs must be different. The replacement of Ga atoms by As atoms is needed to change the sign of $\alpha$. Thus we would have to assume the GaAs crystal consisting of nanocrystals differing from each other by the relative position of Ga and As atoms (antiphase domains [18]). However, the domain boundary would be a source of donors and acceptors that should affect the mobility. In addition, it seems improbable that the same kind of domains would develop in all our GaAs samples grown with different technologies. Further work is necessary to illuminate the origin of the intriguing phenomenon observed here.

To summarize, we have found a giant suppression of electron spin relaxation in high-mobility bulk n-GaAs that persists up to the room temperature. The experimental data obtained with optically oriented electrons can be self-consistently explained by a precession model similar to the Dyakonov-Perel theory, *but with the correlation time of the spin-orbit field being much shorter than, and independent of, the momentum scattering time.* Understanding of this phenomenon may enable high temperature electron spin memory.

The authors are deeply grateful to I.A. Merkulov and V.I. Perel for valuable discussions, I.G. Aksyanov and M.N. Tkachuk for experimental assistance, V.M. Lantratov and S.A. Mintairov for illuminating discussions on antiphase domains, and to S.V. Ivanov and M.N. Stepanova for providing samples. The work has been partially supported by RFBR, CRDF, DARPA/SpinS, NSA/ARDA/ARO, Volkswagen foundation and programs of Russian Academy of Sciences.

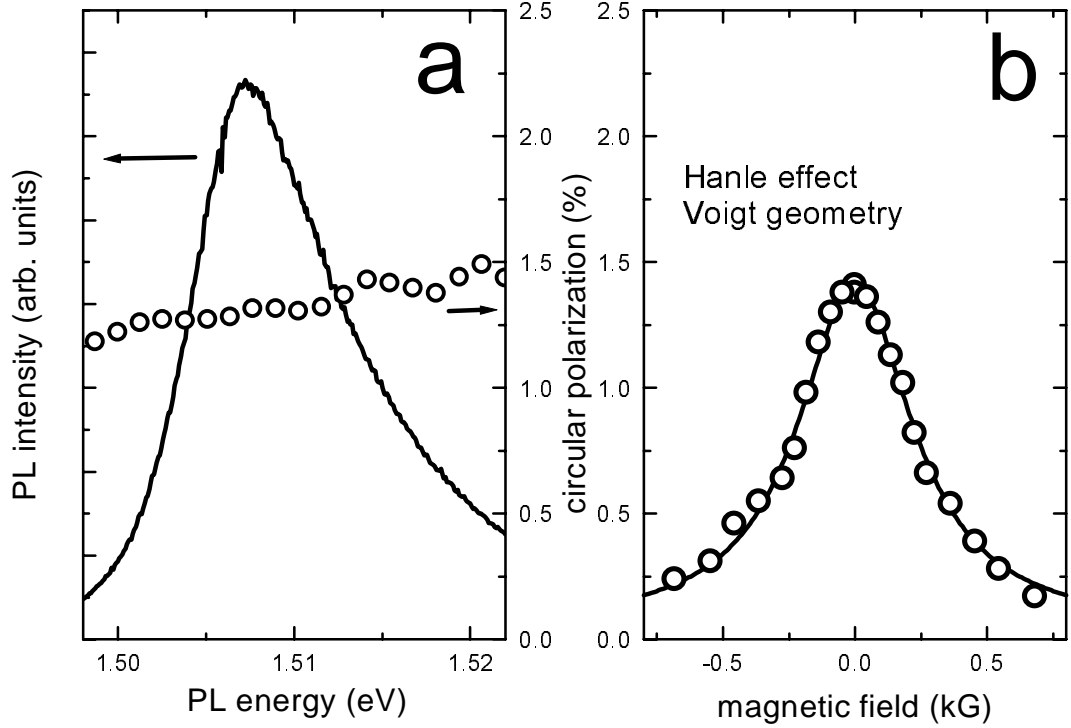

Fig.1. (a): Spectra of PL intensity (solid line) and polarization (circles) at zero magnetic field in a layer of MBE grown GaAs ($N_d-N_a=3*10^{15}$ см$^{-3}$, $\mu_n =29,000$ $cm^2/V*s$), thickness 2$\mu m$, clad with AlGaAs barriers; (b): Hanle effect (circles) measured at the pump density of *1 W/cm$^2$*. Solid line is the Lorentzian fit at *$B_{1/2}=230$ G*, which corresponds to *$1/\tau_s=0.9$ ns$^{-1}$*.

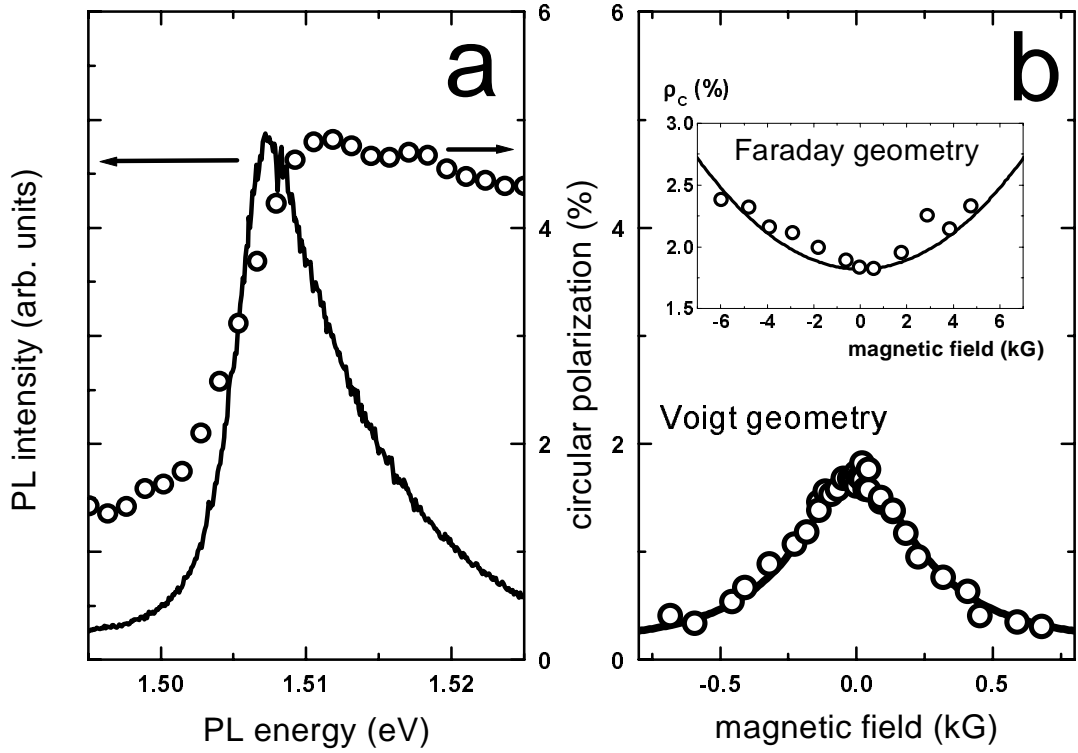

Fig.2. (a): Spectra of PL intensity (solid line) and polarization (circles) at zero magnetic field in a GaAs layer ($N_d-N_a=5*10^{13}$ см$^{-3}$, $\mu_n =140,000$ $cm^2/V*s$), 32$\mu m$ thick, grown by GPE on a semi-insulating GaAs substrate; (b): Hanle effect (circles), excitation density *1.0 W/cm$^2$*. Lorentzian fit (solid curve) gives $B_{1/2}=280$ *G*, corresponding to $1/\tau_s=1.1$ $ns^{-1}$. Inset: dependence of $\rho_c$ on magnetic field in Faraday geometry. Solid curve: fit with Eq.(2) at $\tau_p=270 fs$ .

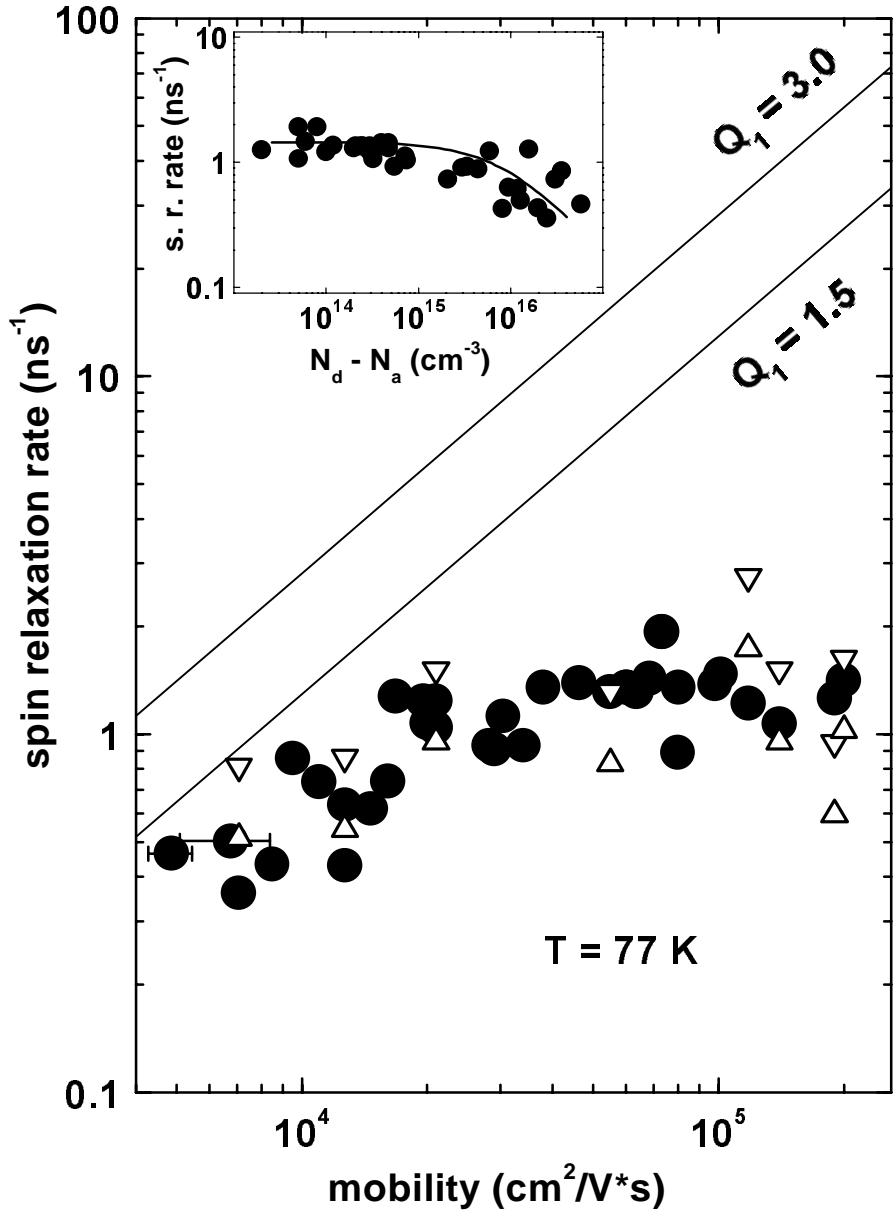

Fig.3. Dependence of the spin relaxation rate of electrons on their mobility. Circles: measurements with the Hanle effect. Solid lines: calculation with Eq.(1) at $Q_1=3$ (scattering by phonons, upper straight line) and $Q_1=1.5$ (scattering by ionized impurities, lower straight line). Triangles: calculation with Eq.(1) using $\tau_p$ measured from spin-relaxation suppression in Faraday geometry, assuming scattering by phonons ($\nabla$), and ionized impurities ($\Delta$). Inset: concentration dependence of the rate $1/\tau_s$. Solid curve is to guide the eye.

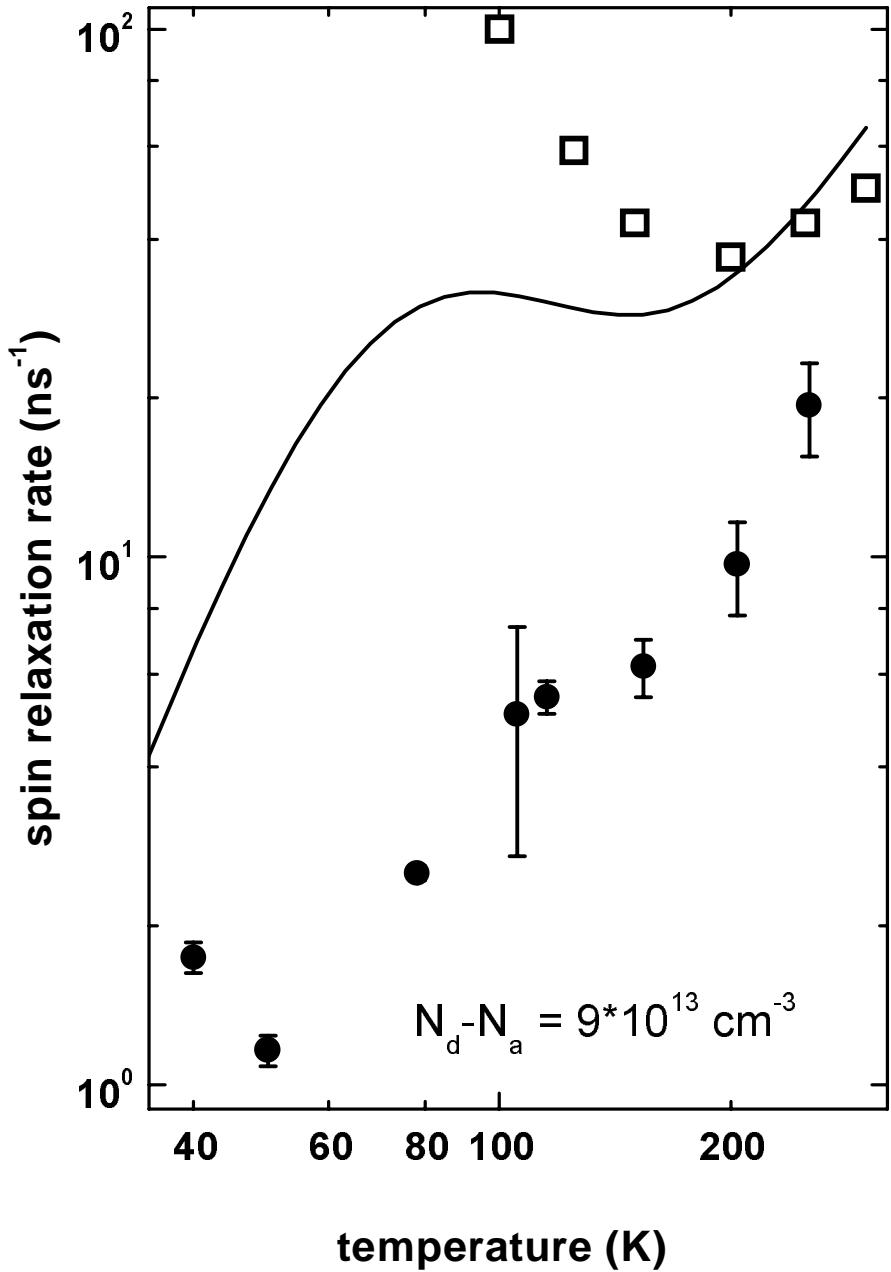

Fig.4. Temperature dependence (solid points) of the rate $1/\tau_s$ of MBE-grown 7.5 μ-thick GaAs with $N_d\text{-}N_a \approx 9*10^{13}$ см$^{-3}$. Solid curve is plotted according to Eq.1 for ionized impurity, acoustic and polar optical phonon scattering elastic processes, whereas the squares are calculated [14] for optical phonon scattering only taking into account it non-elastic character.